# Scattering of Spin-Zero and Spin-Half Particles in Momentum-Helicity Basis

*I. Fachruddin* † *and I. Abdulrahman*

Departemen Fisika, Universitas Indonesia, Depok 16424, Indonesia
† imamf@fisika.ui.ac.id

## Abstract

Scattering of 2 particles of spin 0 and ½ is evaluated based on a basis constructed from the momentum and the helicity states (the momentum-helicity basis). This shortly called three-dimensional (3D) technique is a good alternative to the standard partial wave (PW) technique especially for higher energies, where PW calculations may become not feasible. Taking as input a simple spin-orbit potential model we calculate as an example the spin averaged differential cross section and polarization.

## 1. Introduction

The standard technique used in few-nucleon calculations is the PW technique, which is based on a partial wave projected momentum space basis, see for example Rev. [1], [2]. In this technique only some lowest angular momentum states are taken into account. This is allowed by the short range nature of nuclear interaction. In low energy region taking only a few lowest angular momentum states may adequate, but for higher energies of hundreds MeV states of higher angular momentum must also be considered. This may raise complexities, especially for more complex systems such those of three and four nucleons, which then require intricate numerical calculations as well as tedious algebraical work, see for instance Rev. [3], [4]. Besides in some cases PW calculations seem to almost come to their limits [5].

In Rev. [6] a new technique is developed for nucleon-nucleon (NN) scattering calculation, in which the momentum state is taken without being decomposed into partial waves. The basis state is constructed from the momentum and the helicity states. This so called 3D technique has been successfully applied to calculate NN scattering process based on the NN potentials Bonn-B [7] and AV18 [8].

Now generally the situation described in the first paragraph may also occur in other few-hadron systems. Following the steps given in Rev. [6] we develop a 3D technique for two particles scattering of spin 0 and ½. Thus, given the right potential one can calculate, for example, kaon-nucleon scattering process. Here we take a simple spin-orbit potential to demonstrate the feasibility of the 3D technique.

## 2. Formulation for spin-zero and spin-half particles scattering in momentum-helicity basis

### 2.1. The momentum-helicity basis

Formulation in this work is carried out based on the momentum-helicity basis state defined as:

$$\left|\boldsymbol{p};\hat{\boldsymbol{p}}\tfrac{1}{2}\lambda\right\rangle_\pi = \frac{1}{\sqrt{2}}\left(1+\eta_\pi\mathcal{P}\right)\left|\boldsymbol{p};\hat{\boldsymbol{p}}\tfrac{1}{2}\lambda\right\rangle \qquad (1)$$

with:

$$\left|\boldsymbol{p};\hat{\boldsymbol{p}}\tfrac{1}{2}\lambda\right\rangle = \left|\boldsymbol{p}\right\rangle\left|\hat{\boldsymbol{p}}\tfrac{1}{2}\lambda\right\rangle \qquad (2)$$

In Eqs. (1) & (2) $\boldsymbol{p}$ is the relative momentum between the two particles, $\lambda = \pm\ ½$ the projection of the helicity on the quantization axis $\hat{\boldsymbol{p}}$, $\left|\hat{\boldsymbol{p}}\tfrac{1}{2}\lambda\right\rangle$ the helicity state for spin $s$ = ½, $\mathcal{P}$ the parity operator, $\eta_\pi = \pm1$ the parity eigenvalue and π = ± parity label. Thus, the momentum-helicity basis state has a definite parity.

The normalization of the basis state given in Eq. (1) is obtained as

$$\begin{aligned}&{}_{\pi'}\left\langle\boldsymbol{p}';\hat{\boldsymbol{p}}'\tfrac{1}{2}\lambda'\middle|\boldsymbol{p};\hat{\boldsymbol{p}}\tfrac{1}{2}\lambda\right\rangle_\pi \\ &\qquad = \delta_{\eta_{\pi'}\eta_\pi}\left[\delta\left(\boldsymbol{p}'-\boldsymbol{p}\right)\delta_{\lambda'\lambda} - i\eta_\pi\delta\left(\boldsymbol{p}'+\boldsymbol{p}\right)\delta_{\lambda',-\lambda}\right]\end{aligned} \qquad (3)$$

and the completeness relation results as

$$\sum_{\pi\lambda}\int d\boldsymbol{p}\left|\boldsymbol{p};\hat{\boldsymbol{p}}\tfrac{1}{2}\lambda\right\rangle_\pi \frac{1}{2}\,{}_\pi\left\langle\boldsymbol{p};\hat{\boldsymbol{p}}\tfrac{1}{2}\lambda\right| = 1 \qquad (4)$$

### 2.2. The T-matrix elements and the Lippmann-Schwinger equation

The T-matrix elements in the momentum-helicity basis are defined as

$$T^\pi_{\lambda'\lambda}\left(\boldsymbol{p}',\boldsymbol{p}\right) = {}_\pi\left\langle\boldsymbol{p}';\hat{\boldsymbol{p}}'\tfrac{1}{2}\lambda'\middle|T\middle|\boldsymbol{p};\hat{\boldsymbol{p}}\tfrac{1}{2}\lambda\right\rangle_\pi \qquad (5)$$

The T-matrix elements given in Eq. (5) show some symmetry properties as follow

$$\begin{aligned}T^\pi_{\lambda',-\lambda}\left(\boldsymbol{p}',\boldsymbol{p}\right) &= -i\eta_\pi T^\pi_{\lambda'\lambda}\left(\boldsymbol{p}',-\boldsymbol{p}\right)\\ T^\pi_{-\lambda'\lambda}\left(\boldsymbol{p}',\boldsymbol{p}\right) &= i\eta_\pi T^\pi_{\lambda'\lambda}\left(-\boldsymbol{p}',\boldsymbol{p}\right)\\ T^\pi_{-\lambda',-\lambda}\left(\boldsymbol{p}',\boldsymbol{p}\right) &= T^\pi_{\lambda'\lambda}\left(-\boldsymbol{p}',-\boldsymbol{p}\right)\end{aligned} \qquad (6)$$

Equations (5) & (6) also apply for the potential matrix elements $V^\pi_{\lambda'\lambda}(\boldsymbol{p}',\boldsymbol{p})$. Next, using the symmetry relations given in Eq. (6) leads to the Lippmann-Schwinger equation for the T-matrix elements in the momentum-helicity basis as single – instead of two coupled – integral equation, that is

$$T^{\pi}_{\lambda'\lambda}(\boldsymbol{p}',\boldsymbol{p}) = V^{\pi}_{\lambda'\lambda}(\boldsymbol{p}',\boldsymbol{p}) + \int d\boldsymbol{p}'' V^{\pi}_{\lambda'\frac{1}{2}}(\boldsymbol{p}',\boldsymbol{p}'') G_0^{+}(p'') T^{\pi}_{\frac{1}{2}\lambda}(\boldsymbol{p}'',\boldsymbol{p}) \quad (7)$$

Assuming rotation, parity-operation and time-reversal invariance the potential matrix elements take the general form as

$$V^{\pi}_{\lambda'\lambda}(\boldsymbol{p}',\boldsymbol{p}) = \left[F(p',p,\alpha',\lambda',\lambda) + \eta_{\pi} F(p',p,-\alpha',\lambda',-\lambda)\right]\langle \hat{\boldsymbol{p}}'\lambda' | \hat{\boldsymbol{p}}\lambda \rangle \quad (8)$$

with the $F$ function being determined by the structure of the potential, $\alpha' = \hat{\boldsymbol{p}}' \cdot \hat{\boldsymbol{p}} = cos\theta'$ , the scalar product of the helicity states obtained as

$$\langle \hat{\boldsymbol{p}}'\lambda' | \hat{\boldsymbol{p}}\lambda \rangle = \sum_{m} e^{im(\phi'-\phi)} d^{\frac{1}{2}}_{m\lambda'}(\theta') d^{\frac{1}{2}}_{m\lambda}(\theta) \quad (9)$$

and the rotation matrix given as [9]

$$d^{\frac{1}{2}}_{m'm}(\theta) = \begin{pmatrix} cos\frac{\theta}{2} & -sin\frac{\theta}{2} \\ sin\frac{\theta}{2} & cos\frac{\theta}{2} \end{pmatrix} \quad (10)$$

As usual in calculation for scattering process the direction of the initial momentum is chosen to be in z direction. Using the general form of the potential given in Eq. (8) we obtain the following relation

$$V^{\pi}_{\lambda'\lambda}(\boldsymbol{p}',p\hat{\boldsymbol{z}}) = e^{i\lambda\phi'} V^{\pi}_{\lambda'\lambda}(p',p,cos\theta') \quad (11)$$

which applies also for $T^{\pi}_{\lambda'\lambda}(\boldsymbol{p},p\hat{\boldsymbol{z}})$

$$T^{\pi}_{\lambda'\lambda}(\boldsymbol{p}',p\hat{\boldsymbol{z}}) = e^{i\lambda\phi'} T^{\pi}_{\lambda'\lambda}(p',p,cos\theta') \quad (12)$$

Inserting Eqs. (11) & (12) into Eq. (7) leads to the following integral equation for $T^{\pi}_{\lambda'\lambda}(p',p,\alpha')$

$$T^{\pi}_{\lambda'\lambda}(p',p,cos\theta') = V^{\pi}_{\lambda'\lambda}(p',p,cos\theta') + \int_0^{\infty} dp'' p''^2 \int_{-1}^{1} d(cos\theta'') \mathcal{V}^{\pi\lambda}_{\lambda'\frac{1}{2}}(p',p'',cos\theta',cos\theta'') G_0^{+}(p'') T^{\pi}_{\frac{1}{2}\lambda}(p'',p,cos\theta'') \quad (13)$$

with $\alpha'' = \hat{\boldsymbol{p}}'' \cdot \hat{\boldsymbol{p}} = cos\theta''$ and

$$\mathcal{V}^{\pi\lambda}_{\lambda'\frac{1}{2}}(p',p'',cos\theta',cos\theta'') = \int_0^{2\pi} d\phi'' e^{i\lambda(\phi''-\phi')} V^{\pi}_{\lambda'\frac{1}{2}}(\boldsymbol{p}',\boldsymbol{p}'') \quad (14)$$

We solve Eq. (13) to obtain $T^{\pi}_{\lambda'\lambda}(p',p,\alpha')$ and use $T^{\pi}_{\lambda'\lambda}(p',p,\alpha')$ as input to calculate scattering observables.

Fortunately, the T-matrix elements $T^{\pi}_{\lambda'\lambda}(p',p,\alpha')$ for various values of $\lambda'$ and $\lambda$ are connected to each other as shown by the following symmetry relation

$$\begin{aligned} T^{\pi}_{-\lambda'\lambda}(p',p,cos\theta') &= (-)^{\lambda} i\eta_{\pi} T^{\pi}_{\lambda'\lambda}(p',p,-cos\theta') \\ T^{\pi}_{\lambda',-\lambda}(p',p,cos\theta') &= (-)^{\lambda'} i\eta_{\pi} T^{\pi}_{\lambda'\lambda}(p',p,-cos\theta') \\ T^{\pi}_{-\lambda',-\lambda}(p',p,cos\theta') &= -T^{\pi}_{\lambda'\lambda}(p',p,cos\theta') \end{aligned} \quad (15)$$

Therefore, we need to solve only one equation for each parity.

### 2.3. The potential as input

The input for the calculations is the potential operator $V$. For the basis state given in Eq. (1) the convenience form of the potential matrix elements in momentum space is a function of the spin operators $\boldsymbol{\sigma} \cdot \hat{\boldsymbol{p}}'$ and $\boldsymbol{\sigma} \cdot \hat{\boldsymbol{p}}$ as well as their combination, with $\boldsymbol{\sigma} \cdot \hat{\boldsymbol{p}}'$ appearing to the left of $\boldsymbol{\sigma} \cdot \hat{\boldsymbol{p}}$ , thus,

$$V(\boldsymbol{p}',\boldsymbol{p}) = \langle \boldsymbol{p}' | V | \boldsymbol{p} \rangle = \sum_{i} f_i(p',p,\hat{\boldsymbol{p}}' \cdot \hat{\boldsymbol{p}}) (\boldsymbol{\sigma} \cdot \hat{\boldsymbol{p}}')^{a_i} (\boldsymbol{\sigma} \cdot \hat{\boldsymbol{p}})^{b_i} \quad (16)$$

In Eq. (16) $a_i$ and $b_i$ have to be such that the potential meets the requirement rotation, parity-operation and time-reversal invariance. Such potential will lead to the potential matrix elements given in Eq. (8).

Any potential given in different expression from the one in Eq. (16) have to be reexpressed, so that it has the form given in Eq. (16). This is performed by relating spin operators appearing in the potential to $\boldsymbol{\sigma} \cdot \hat{\boldsymbol{p}}'$ and $\boldsymbol{\sigma} \cdot \hat{\boldsymbol{p}}$ .

As an example we take the following simple spin-orbit potential

$$V(r) = V_c(r) + V_s(r)\boldsymbol{l} \cdot \boldsymbol{s} \quad (17)$$

with $\boldsymbol{l} = \boldsymbol{r} \times \boldsymbol{p}$ and $\boldsymbol{s} = \frac{1}{2}\boldsymbol{\sigma}$ . In momentum space the potential given in Eq. (17) is obtained as

$$V(\boldsymbol{p}',\boldsymbol{p}) = V_c(\boldsymbol{p}',\boldsymbol{p}) + V_s(\boldsymbol{p}',\boldsymbol{p})\boldsymbol{s} \cdot (\boldsymbol{p} \times \boldsymbol{p}') \quad (18)$$

It can be shown that the potential given in Eq. (17) or (18) is invariant under rotation, parity operation and time reversal. The potential in Eq. (18) can be reexpressed to have the form given in Eq. (16) using the following relation

$$\boldsymbol{s} \cdot (\boldsymbol{p} \times \boldsymbol{p}') = -\frac{i}{2} p'p \left\{ \hat{\boldsymbol{p}}' \cdot \hat{\boldsymbol{p}} - (\boldsymbol{\sigma} \cdot \hat{\boldsymbol{p}}')(\boldsymbol{\sigma} \cdot \hat{\boldsymbol{p}}) \right\} \quad (19)$$

## 3. Example of calculation: the spin averaged differential cross section and polarization

Finally in this section we show as an example some results from our of calculations using the 3D technique. We choose a process as if it is kaon-nucleon scattering, where the nucleon acts as the projectile and the scattered particle. We calculate the spin averaged differential cross section and polarization of the scattered nucleon. As input we take the potential given in Eq. (17), with $V_c(\boldsymbol{r})$ and $V_s(\boldsymbol{r})$ take the form of the Malfliet-Tjon potential [10]:

$$V_c(r) = -V_{ca}\frac{e^{-\mu_c r}}{r} + V_{cb}\frac{e^{-2\mu_c r}}{r}$$
$$V_s(r) = -V_{sa}\frac{e^{-\mu_s r}}{r} + V_{sb}\frac{e^{-2\mu_s r}}{r} \quad (20)$$

with the parameters being chosen without any preference as

$$V_{ca} = 3.22 \quad V_{cb} = 7.39 \quad \mu_c = 1.55\ fm^{-1}$$
$$V_{sa} = 2.64 \quad V_{sb} = 7.39 \quad \mu_s = 0.63\ fm^{-1} \quad (21)$$

The spin averaged differential cross section for various nucleon's laboratorium energies is displayed in Fig. 1 and polarization of the scattered nucleon in Fig. 2.

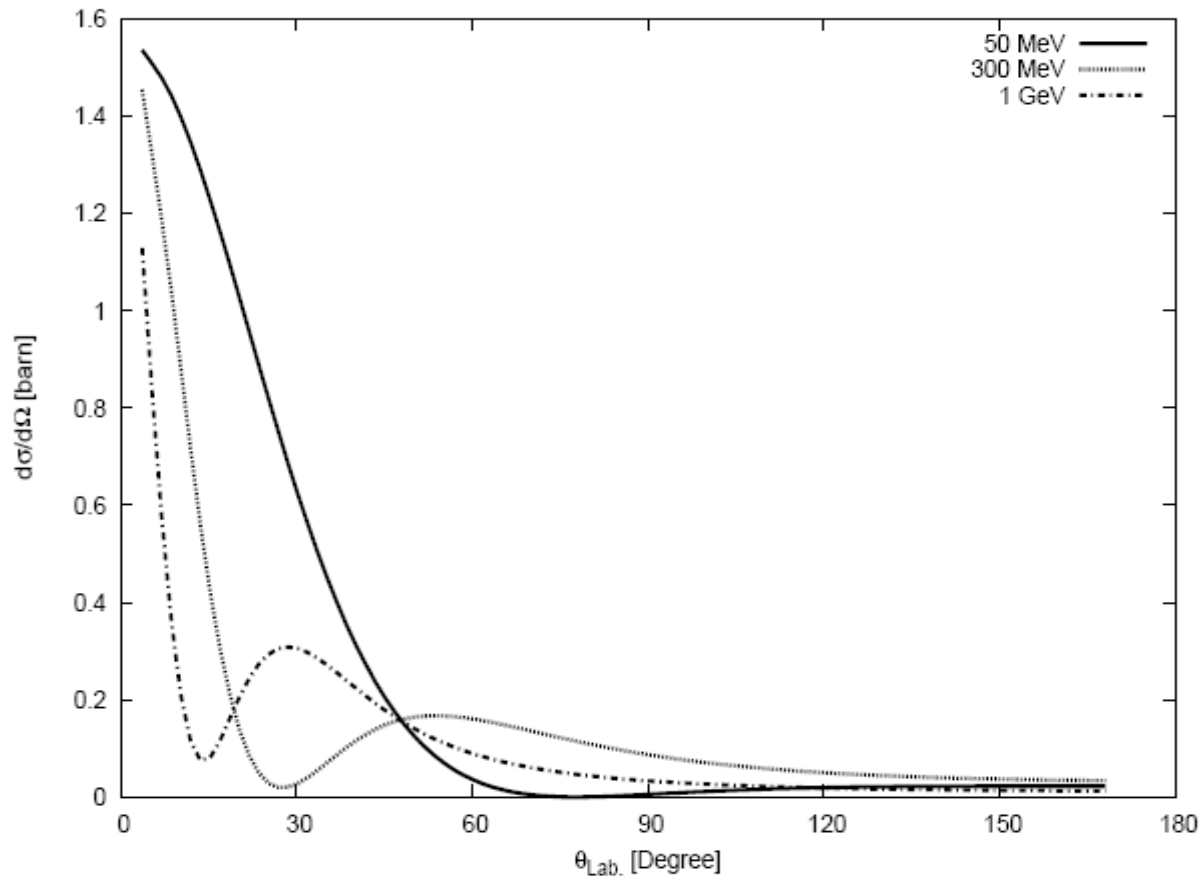

*Figure 1:* The spin averaged differential cross section for various nucleon's laboratorium energies

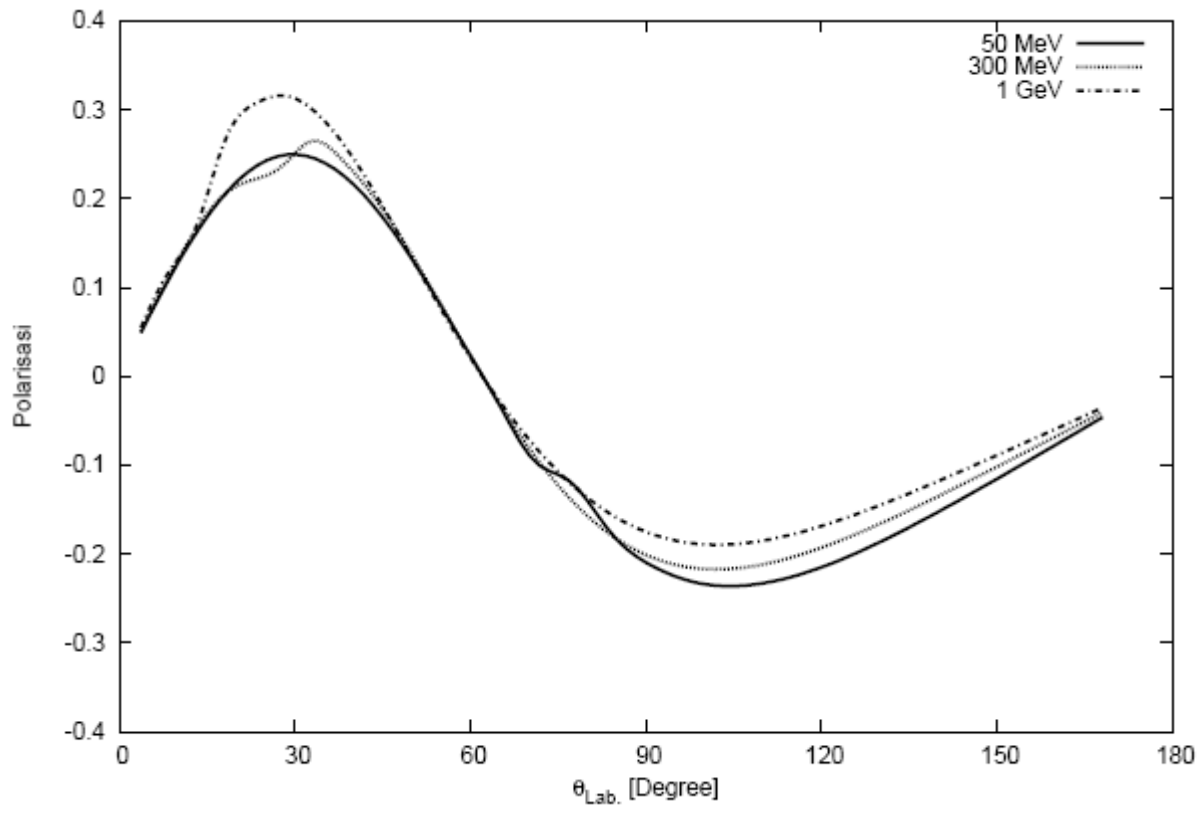

*Figure 2:* Polarization of the scattered nucleon for various nucleon's laboratorium energies

The calculations have been performed using a PC and it takes only a few minutes for nowadays processors installed on PC's. Of course if the potential is more complicated it takes more time to finish the calculation. As described in Section 2 using the 3D technique one needs only to solve a fix number of equation (two equations) regardless scattering energies. Hence, it takes the same effort to calculate the scattering process for high energy as that for low energy. This is not the case in PW calculations, where the higher the energy the more effort required.

## 4. Summary

We have develop a technique – called a 3D technique – to calculate scattering of two particles of spin 0 and ½. The technique has shown to be a good alternative to the PW technique, especially for higher energy of a few hundreds MeV, where PW calculations may become not feasible. As an example we have performed some calculations based on a simple spin-orbit potential model.